# T2-weighted Spine Imaging using a Single-Shot Turbo Spin Echo Pulse Sequence

Mahesh Bharath Keerthivasan, MS,[1,2], Blair Winegar, MD,[2] Jennifer L Becker, MD,[2] and Manojkumar Saranathan, PhD,[2]

1 Department of Electrical and Computer Engineering, University of Arizona, Tucson, Arizona, USA
2 Department of Medical Imaging, University of Arizona, Tucson, Arizona, USA
3 Department of Biomedical Engineering, University of Arizona, Tucson, Arizona, USA

## ABSTRACT

**Objectives:**

To investigate the use of a half-Fourier acquisition single-shot turbo spin-echo variable flip angle (HASTE-VFA) sequence for T2 weighted spine imaging.

**Materials and Methods:**

T2 weighted imaging of the spine is commonly performed using fast spin echo (FSE/TSE) based sequences, resulting in long scan times and vulnerability to motion artifacts. While single shot fast spin echo sequences have been attempted, their adoption has been limited by poor spatial resolution and specific absorption rate (SAR) limitations. A variable refocusing flip angle echo train was first optimized for the spine to improve the point spread function (PSF) and minimize SAR, yielding images with improved spatial resolution and signal-to-noise ratio compared to the constant flip angle sequence. Data was acquired from 29 patients (20 lumbar and thoracolumbar, 9 whole-spine) using conventional fast spin echo and the proposed variable flip angle single-shot sequences. All images were graded by two experienced neuroradiologists in a blinded fashion and scores were assigned based on blurring, motion, artifacts, and noise as well as appearance of disks, facet joints, end plates, nerve roots and the spinal cord.

**Results:**

The HASTE-VFA sequence had a 3–4x reduction in scan time compared to the TSE. The mean scores for the lumbar and thoracolumbar cases were higher for the TSE compared to the proposed HASTE-VFA sequence. However, HASTE-VFA scored >=4 for all the criteria indicating its diagnostic potential. The mean scores for the whole spine cases for the HASTE-VFA sequence were higher than that of the lumbar and thoracolumbar cases.

**Conclusion:**

We present a fast and motion robust T2 weighted spine protocol using the variable flip angle HASTE sequence. The sequence has better PSF behavior than the constant flip angle variant and being a quick scan it is insensitive to patient motion, often seen in the longer TSE scans. This would enable the use of the sequence in spine screening protocols.

**Keywords:**

Single shot fast spin echo, variable flip angle, spine imaging, T2 weighted MRI

**INTRODUCTION**

T2-weighted (T2w) magnetic resonance imaging (MRI) protocols are routinely used in the clinic for spine imaging. T2w sequences such as 2D fast spin echo (FSE) or turbo spin echo (TSE) have proved useful for detection and diagnosis of osteomyelitis, stenosis, nerve root compression, disk characterization and other pathology[1,2,3]. A typical spine protocol includes sequences for T1 weighted, T2 weighted and short-time inversion recovery (STIR) fat suppressed T1 imaging, all with FSE readouts. The 2D T2 weighted sequences have long acquisition times of about 3-4 minutes and these times double or triple when performing combination spine exams (e.g. thoracolumbar or whole spine). This makes the scan uncomfortable for patients with painful spine conditions and also makes the imaging more vulnerable to motion artifacts. Approaches using parallel imaging[4] have been explored to reduce the scan time, however the use of higher acceleration factors is restricted by the image quality degradation arising from poor coil coverage.

The 2D single shot fast spin echo sequence (SSFSE) or half-Fourier acquisition single-shot turbo spin-echo sequence (HASTE) has been shown to be useful for applications such as lumbar spine myelography[5], cervical spine motion imaging[6] and functional imaging of the spine[7]. However, the sequence is not preferred for routine T2w imaging since it suffers from image blurring and specific absorption rate (SAR) limitations arising from the long echo train, often addressed by scanning at low spatial resolution[8]. Despite these compromises, HASTE is highly SAR limited at 3T field strength.

3D fast spin echo sequences with very long echo train lengths that use variable flip angle (VFA) schemes for the refocusing train have been very successful in a number of applications especially brain imaging[9,10,11]. This approach allows optimization of the refocusing flip angle train to improve the imaging point spread function, leading to reduced T2 blurring and improved signal-to-noise ratio (SNR) and SAR performance. The variable flip angle technique has very recently been applied to 2D single shot fast spin echo sequences[12] and was successfully demonstrated for body applications [13,14] with significant reduction in scan times due to the reduced SAR and increased sharpness compared to conventional SSFSE.

In this work, we investigated the use of a variable flip angle single shot HASTE sequence (HASTE-VFA) optimized for T2 weighted spine imaging. We explored the use of this fast motion robust sequence in a clinical setting for two purposes - (a) as a potential fast adjunct sequence to conventional TSE to reduce scan time in patients who cannot tolerate long examinations and (b) as a quick T2w screening sequence with diagnostic image quality in fast imaging protocols where the use of TSE is prohibitively long (e.g. whole spine protocol for trauma).

**MATERIALS AND METHODS**

**Pulse Sequence and Parameter Optimization**

A commercial HASTE sequence was modified to incorporate the refocusing flip angle modulation scheme proposed by Busse et al[9]. This technique parameterizes the refocusing flip angle train using four control angles ($\alpha_{start}, \alpha_{min}, \alpha_{cent}, \alpha_{end}$). Following the approach outlined by Loening et al[14], the flip angles were optimized to maximize the SNR for the desired effective echo time (TE) and subsequently minimize SAR with minimal spatial blurring. We specifically used T1 and T2 values for the spine at 3T (T1=1060 ms,

T2=69 ms) and MATLAB simulations using Extended Phase Graph (EPG) simulations to compute optimal refocusing flip angle trains for spine imaging. The reduction in SAR is especially important at 3T as it reduces the scan time for fast spin echo based sequences which have longer SAR limited repetition times (TR).

The HASTE-VFA sequence was implemented and tested on a Siemens 3T (Skyra, Siemens, Erlangen, Germany) scanner. The HASTE-VFA protocol was first optimized on healthy volunteers after informed consent. Data was acquired using the constant flip angle HASTE and the HASTE-VFA sequences to study the effect of blurring and to optimize the sequence parameters, using the MATLAB simulation based parameters as a starting point. Phase encoding gradients along the anterior-posterior (AP) direction and readout flow compensation to reduce CSF pulsation artifacts were employed. Note that in conventional TSE, the frequency encoding gradients are applied along the AP direction to minimize motion artifacts from breathing, swallowing etc. The use of an ultrafast HASTE-VFA (~680 ms per slice) sequence allowed us use to make this switch enabling more optimal spatial resolution as the AP direction is usually much narrower than the SI direction in the case of spine imaging. Since the T2 decay effects are prolonged and SAR is minimized in variable flip angle sequences, a partial Fourier factor of 6/8 could be used to achieve a desired effective TE of 108 msec. A slice concatenation factor of 3 or 4 was used to reduce inter-slice cross talk artifacts.

**Clinical Imaging**

Written consent was obtained from subjects in compliance with the Institutional Review Board requirements. The optimized HASTE-VFA sequence was then added to the lumbar spine, the thoracic spine and the thoracolumbar clinical spine protocols as an addition to the routine T2 weighted TSE sequence. It was also added to the whole spine protocol (covering the cervical, thoracic and lumbar spine) which omits the TSE sequence due to its long scan time (~9 min). Since the constant flip angle HASTE sequence showed severe blurring in the volunteer scans, it was not added to the clinical protocol. Data were acquired from 29 consecutive patients scanned between November 19, 2016 and January 10, 2017. Of these 9 were whole spine exams, 10 lumbar and the remaining thoracolumbar exams.

The scan parameters of the TSE and the HASTE-VFA sequences are shown in Table 1 for the different spine protocols. The effective TE was set to 108 ms for both the sequences and images were acquired in the sagittal plane. Note that the acquisition spatial resolution achieved by the HASTE-VFA sequence is less than that of the TSE sequence, as it is ultimately bound by SNR considerations. For the constant flip angle HASTE sequence most of the parameters were the same as that of the HASTE-VFA.

**Image Quality Assessment**

The performance of the HASTE-VFA and the TSE sequences were assessed independently by two neuroradiologists specialized in spine MRI in a blinded fashion. In order to avoid recollection bias, images from the two sequences were scored one week apart with random assignment of one or the other sequence to each reading. Images were graded on a scale of 1-5 (1 - non-diagnostic images, 2 – severely limited, 3 – limited, 4 – one or two suboptimal attributes but still diagnostic and 5 – optimal image quality for diagnosis). Images were graded based on the following criteria: edge sharpness, motion, artifacts and noise. Clinical utility of the sequences was assessed by quantifying the ability of the reader to interpret facet joints, endplates, nerve roots, spinal cord, and discs, also on a scale of 1-5.

**Statistical Analysis**

Data from the lumbar and thoracolumbar spine cases were pooled and a Wilcoxon signed rank test with a p-value of 0.05 was performed with the null hypothesis that the image quality of the two sequences do not have any significant differences. Cohen's Kappa statistic was initially used to assess inter-observer reliability. However, it was observed that the statistic yielded low values for grading criteria that have a high percentage of agreement and are skewed in the score distribution. In order to overcome this so-called Kappa paradox[15], Gwet's AC1 statistic was instead used to measure inter-observer variability as it has been shown[16] to be more robust to skewed distributions. Since the whole spine exams did not have a TSE reference, their scores were analyzed separately.

**RESULTS**

Figure 1 illustrates a flip angle modulation scheme and compares the T2 signal decay for a constant and a variable flip angle echo train. The VFA scheme stabilizes the signal evolution over the echo train and improves the effective area under the T2 decay curve (SNR) when compared to the constant flip angle. Point spread function (PSF) analysis of the two sequence variants was performed to evaluate the spatial blurring behavior. Since the constant flip angle sequence is typically acquired with partial Fourier, the PSF was computed using a projection onto convex sets (POCS) reconstruction model. From the plots in Figure 1C the improvement in the spatial resolution with the variable refocusing flip angle train can be clearly observed, albeit at the expense of a slight reduction in signal

In order to verify the improvement in PSF behavior, volunteer data was acquired using a constant flip angle HASTE sequence and a HASTE-VFA sequence. A T2 weighted TSE sequence was also acquired as reference. As seen in Figure 2, the image from HASTE-VFA has significantly reduced blurring and has comparable contrast to the TSE sequence (effective TE=105ms).

Despite the use of refocusing flip angles of $120^0$, the constant flip angle HASTE had a SAR limited TR value of 1700 ms compared to 700 ms for the HASTE-VFA sequence (Table 1). The SAR for the variable flip angle sequence was 1.2 times lower than the constant flip angle sequence and 1.6 times lower than the TSE sequence.

Figure 3 illustrates the effect of subject motion on a patient with discitis and osteomyelitis. The thoracic spine image exhibits motion artifacts in the TSE image, while the HASTE-VFA generates image with comparable contrast and reduced motion artifacts due to its reduced scan time. The increased blurring from the reduced spatial resolution of the HASTE-VFA sequence relative to the TSE sequence can also be seen.

Representative examples of lumbar and thoracolumbar spine images acquired on patients using the TSE and the HASTE-VFA sequences are shown in Figures 4 and 5. Note that the variable flip angle sequence generates images of comparable contrast, albeit at a slightly lower resolution when compared to the TSE.

Figures 6 and 7 illustrate the utility of the fast HASTE-VFA sequence in a whole spine protocol acquired in less than 2 minutes. Due to its long scan time (9 min for all three stations) the TSE sequence was replaced by the HASTE-VFA sequence for T2 weighted imaging. The image quality and spatial resolution are comparable to the STIR sequence (9 min scan time).

Table 2 summarizes the mean scores from the two observers for TSE and HASTE-VFA for the lumbar and thoracolumbar spine exams. The TSE scores were significantly higher than HASTE-VFA for all the criteria ($p < 0.05$). However, the mean scores for HASTE-VFA were still >= 4 (except for noise which was 3.9) indicating that the sequence is still diagnostically useful, despite the lower resolution. The scores for the discs and endplates were higher than that of the spinal cord and facet joints for the HASTE-VFA. The inter-observer reliability AC1 scores and the percentage agreement between readers are listed in Table 3. It can be seen that there was a good level of agreement between the readers for all the criteria scored (except for edge sharpness in the HASTE-VFA), although the agreement was slightly higher for TSE than for HASTE-VFA.

The mean scores for the whole spines along with Gwet's AC1 and percentage agreement values are shown in Table 4. It can be seen that there is good inter-observer concordance in the scores except for artifacts and noise. It can also be noted that the mean HASTE-VFA image scores were either similar or slightly higher than for the lumbar/thoracic case. This is presumably due to the nature of the whole spine protocol which is used more for screening than for looking at detailed structures.

**DISCUSSION**

We have demonstrated the use of single-shot fast spin echo sequences with variable refocusing flip angles as a fast and motion robust adjunct for routine spine imaging and a valuable addition for whole spine imaging protocols where TSE sequences are prohibitively long. The use of flip angle modulation reduced the blurring and SAR typically associated with single shot FSE sequences. The refocusing flip angles of the HASTE-VFA sequence were optimized to maximize the spatial resolution and minimize the SAR. However, the overall resolution was still limited by the available signal to noise ratio (SNR), resulting in a lower resolution than the TSE (0.81 x 0.73 mm vs 1.25 x 1.0 mm). This was presumably the reason for lower image quality scores compared to the TSE sequence. However, the scores were still 4 or higher for the overall imaging and the clinical criteria used, implying that HASTE-VFA can still provide images with diagnostic quality, albeit with some limitations like reduced spatial resolution and/or SNR. Preliminary studies on volunteers indicates that the use of anterior coil arrays (which is atypical in spine imaging) in addition to the posterior array would significantly improve SNR and enable comparable spatial resolution between HASTE-VFA and TSE. We are currently exploring the feasibility of this in the clinical setting.

The HASTE-VFA sequence was added to the whole spine protocol as a fast T2w screening scan, and its diagnostic value was confirmed by the excellent image quality scores (the diagnostic criteria had an average score of 4.4 and the image quality metrics showed better artifact behavior and SNR compared to the lumbar and thoracic spine). For the lumbar / thoracolumbar spine protocol, the HASTE-VFA sequence had an acquisition time of 1.03 minutes making it less vulnerable to patient motion compared to the 3.63 minute TSE scan. This makes HASTE-VFA a good candidate sequence that can be added to the spine protocol along with the T2w TSE for routine spine imaging where it would help in the diagnosis of cases where TSE is corrupted by motion artifacts or in cases where the patient is unable to tolerate long scans due to physical limitations like pain.

The resolution of the HASTE-VFA was observed to be suboptimal in cervical spine cases. This could be possibly attributed to the inherently poor SNR and artifacts arising from periodic motion such as swallowing which are typically worst in the case of cervical spine imaging compared to other spine imaging protocols. The signal dropouts observed in the body imaging applications[14] with this sequence were not observed in the spine exams.

The proposed sequence cannot be used for T1 weighted imaging due to limitations on the minimum possible effective TE that can be achieved. This could be overcome with the use of non-Cartesian acquisition trajectories (e.g. radial) and specialized reconstruction schemes[17] that can be used to reconstruct the individual echoes. The sequence also needs to be evaluated with a STIR preparation scheme to evaluate its efficiency as a fat suppressed T2w sequence, as well as for imaging in the axial plane where scan times are even longer for contiguous total spine coverage.

In conclusion, we have developed a fast and motion robust T2 weighted spine protocol using the variable flip angle HASTE sequence. The sequence has better PSF behavior than the constant flip angle variant and, being a quick scan, is insensitive to patient motion, often seen in the longer TSE scans. This would enable the use of the sequence as an adjunct or replacement in clinical spine screening protocols and for pediatric 3T spine protocols where TSE scans are highly limited by SAR.

**FIGURES**

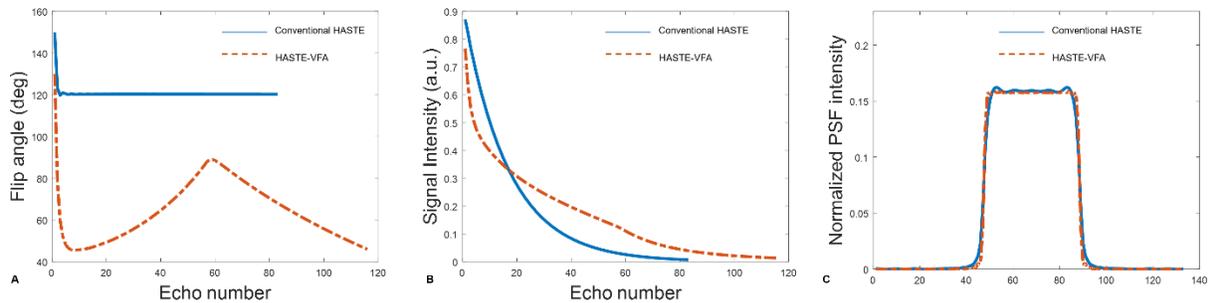

Figure 1. (A) shows the flip angle modulation scheme for a conventional HASTE and the variable flip angle HASTE sequence along with the T2 signal evolution (B). The point spread functions (PSF) for the constant and the variable flip angle echo trains are compared in (C). Note that there is a significant improvement in the PSF with the use of variable refocusing flip angles, resulting in better spatial resolution and less blurring.

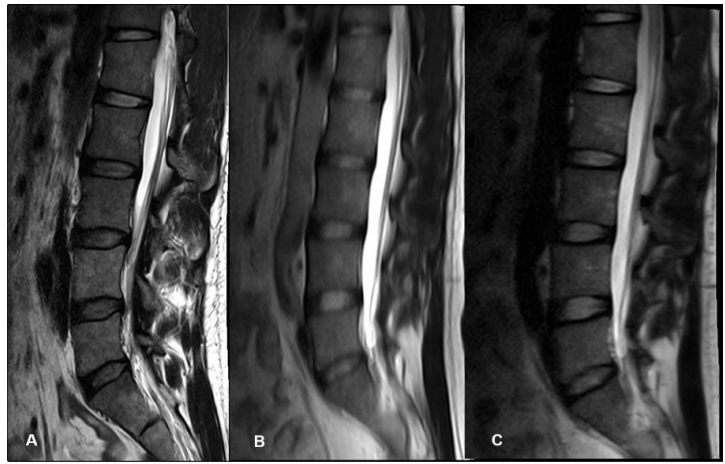

Figure 2. Better PSF behavior (reduced blurring) with variable flip angle HASTE (C ) compared to constant flip angle HASTE (B). TR was reduced from 1700 ms to 700 ms due to the reduction in specific absorption rate. Note that at the effective TE=105 ms the contrast between HASTE-VFA and T2-TSE sequence (A) is comparable.

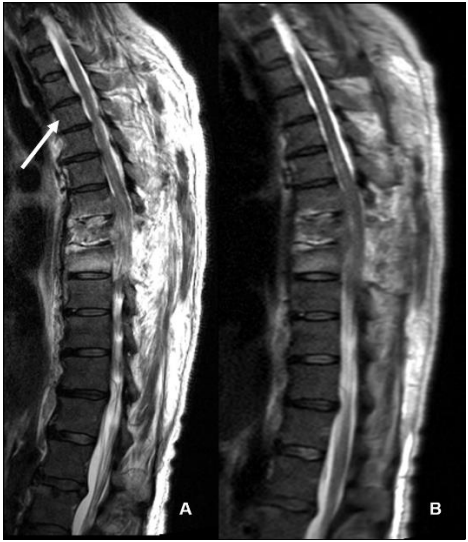

Figure 3. T2 weighted thoracic spine image acquired using the TSE sequence (A) shows motion artifacts (arrow). The HASTE-VFA image (B) has comparable contrast and both the sequences show discitis and osteomyelitis.

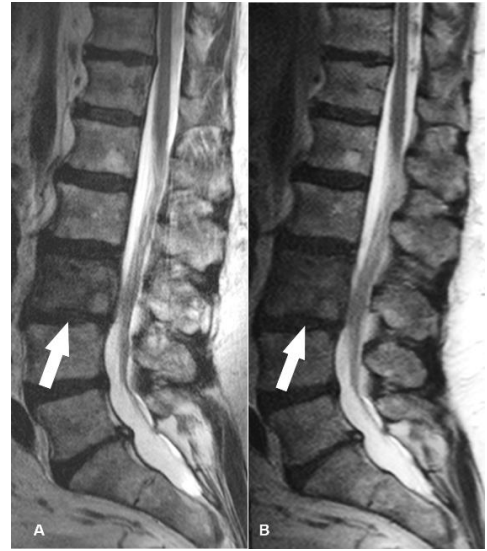

Figure 4. Sagittal T2 TSE (A) and sagittal HASTE-VFA (B) of the lumbar spine demonstrate low signal mass filling the L3 vertebral body (arrows) in a patient with prostate cancer.

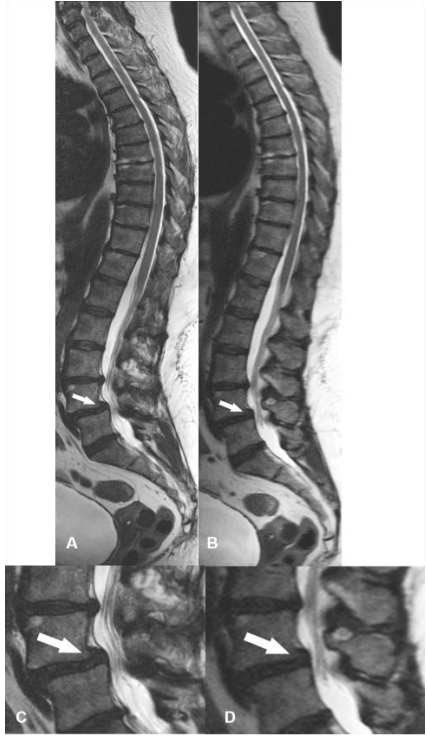 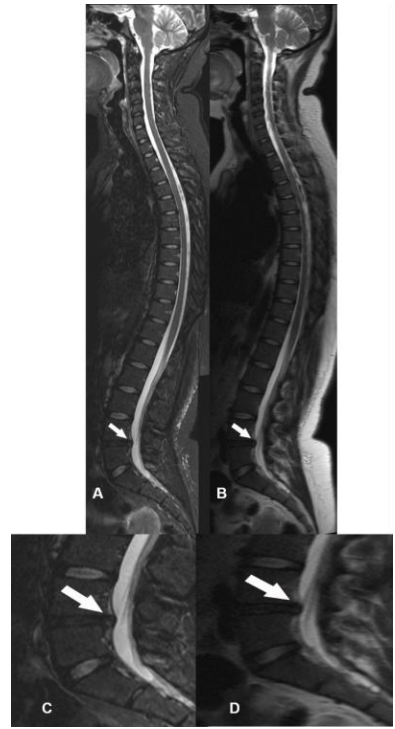

Figure 5. Sagittal T2 TSE (A) and sagittal HASTE-VFA (A) of the thoracolumbar spine demonstrate grade I anterolisthesis of L4 on L5 (arrows) with associated disc bulge resulting in moderate spinal canal stenosis. (C) and (D) show the region of pathology for the TSE and the HASTE-VFA respectively.

Figure 6. Sagittal STIR (A) and sagittal HASTE-VFA (A) of the whole spine demonstrate low signal of the L4-L5 intervertebral disc with disc bulge resulting in mild spinal canal stenosis (arrows). The region of interest for the TSE and the VFA sequence are highlighted in inserts (C) and (D) respectively. The HASTE-VFA (2 min scan) exhibits comparable image quality as the longer STIR scan (9 min).

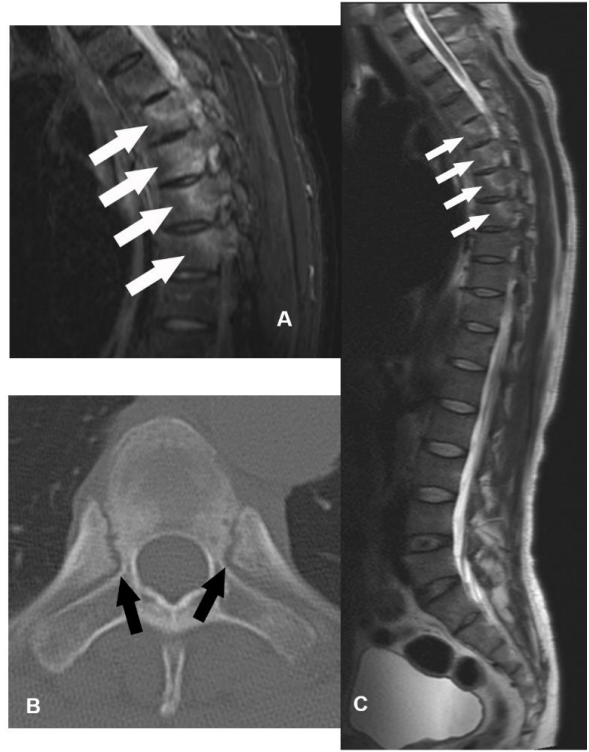

Figure 7. Sagittal STIR of the thoracic spine (A) and sagittal HASTE-VFA of the thoracolumbar spine (C) demonstrate high signal adjacent to multiple upper thoracic costovertebral joints (white arrows) in the patient with seronegative spondyloarthropathy. Axial bone algorithm CT of the thoracic spine (B) demonstrates subchondral sclerosis and erosions (black arrows) of the costovertebral joints.

Table 1: Scan parameters for the different spine protocols at 3T

| Parameters | Lumbar Spine | | Thoracic Spine | | Thoracolumbar spine | | Whole Spine |
|---|---|---|---|---|---|---|---|
| | TSE | HASTE-VFA | TSE | HASTE-VFA | TSE | HASTE-VFA | HASTE-VFA |
| Acquisition resolution (phase x frequency) (mm$^2$) | 0.81 x 0.73 | 1.26 x 1.04 | 0.81 x .73 | 1.14 x 1.0 | 0.81 x 0.73 | 1.25 x 1.0 | 1.25 x 1.0 |
| Reconstruction resolution (phase x frequency) (mm2) | 0.36 x 0.36 | 0.52 x 0.52 | 0.73 x 0.73 | 0.5 x 0.5 | 0.36 x 0.36 | 0.5 x 0.5 | 0.5 x 0.5 |
| Slice Thickness (mm) | 3 | 3 | 3 | 3 | 3 | 3 | 3 |
| Refocusing Flip Angle | 140$^0$ | $\alpha_{start} = 130°$ $\alpha_{min} = 45°$ $\alpha_{cent} = 90°$ $\alpha_{end} = 45°$ | 140° | $\alpha_{start} = 130°$ $\alpha_{min} = 50°$ $\alpha_{cent} = 90°$ $\alpha_{end} = 45°$ | 140° | $\alpha_{start} = 130°$ $\alpha_{min} = 50°$ $\alpha_{cent} = 90°$ $\alpha_{end} = 45°$ | $\alpha_{start} = 130°$ $\alpha_{min} = 50°$ $\alpha_{cent} = 90°$ $\alpha_{end} = 45°$ |
| Averages | 2 | 3 | 2 | 3 | 2 | 3 | 3 |
| Parallel Imaging acceleration factor | 2 | -- | 2 | -- | 2 | -- | -- |
| Partial Fourier factor | -- | 7/8 | -- | 6/8 | -- | 6/8 | 6/8 |
| Phase Encoding direction | Superior-Inferior | Anterior-Posterior | Superior-Inferior | Anterior-Posterior | Superior-Inferior | Anterior-Posterior | Anterior-Posterior |
| TR (ms) | 2800 | 615 | 2870 | 772 | 2800 | 772 | 772 |
| Scan time (min) | 3.63 | 1.03 | 3.73 | 0.76 | 7.27 | 1.47 | 1.95 |

Table 2: Image quality assessment scores for the lumbar and thoracolumbar cases. The mean scores with the standard deviation are shown for the two sequences and the criteria that yielded p-values < 0.05 in the Wilcoxon test are indicated by an asterisk (*).

| Scoring Criteria | T2- TSE Mean Score | HASTE-VFA Mean Score |
|---|---|---|
| Motion | 4.8 ± 0.44 | 4.8 ± 0.41 |
| Edge sharpness | 4.6 ± 0.6 | 4.2 ± 0.83 * |
| Artifacts | 4.9 ± 0.33 | 4.2 ± 0.5 * |
| Noise | 4.8 ± 0.42 | 3.9 ± 0.65 * |
| Facet joints | 4.9 ± 0.33 | 4.2 ± 0.53 * |
| Endplates | 4.9 ± 0.33 | 4.4 ± 0.53 * |
| Spinal cord | 4.8 ± 0.42 | 4 ± 0.53 * |
| Discs | 4.6 ± 0.64 | 3.9 ± 0.6 * |

Table 3: Inter-observer reliability based on Gwet's AC1 for the lumbar and thoracolumbar ratings.

| Scoring Criteria | TSE | | | | HASTE-VFA | | | |
|---|---|---|---|---|---|---|---|---|
| | Reviewer 1 Mean Score | Reviewer 2 Mean Score | Gwet's AC1 | % Agreement | Reviewer 1 Mean Score | Reviewer 2 Mean Score | Gwet's AC1 | % Agreement |
| Motion | 5 ± 0 | 4.5 ± 0.51 | 0.45 | 50 | 4.9 ± 0.22 | 4.7 ± 0.49 | 0.67 | 70 |
| Edge sharpness | 4.8 ± 0.55 | 4.4 ± 0.59 | 0.54 | 60 | 4.5 ± 0.83 | 3.8 ± 0.7 | 0.17 | 30 |
| Artifacts | 4.9 ± 0.31 | 4.8 ± 0.37 | 0.74 | 75 | 4.35 ± 0.55 | 4.1 ± 0.44 | 0.61 | 65 |
| Noise | 4.8 ± 0.41 | 4.8 ± 0.44 | 0.84 | 85 | 4.1 ± 0.64 | 3.7 ± 0.59 | 0.48 | 55 |
| Facet joints | 4.9 ± 0.31 | 4.8 ± 0.37 | 0.74 | 75 | 4.1 ± 0.56 | 4.2 ± 0.52 | 0.55 | 60 |
| Endplates | 4.9 ± 0.22 | 4.8 ± 0.41 | 0.74 | 75 | 4.4 ± 0.5 | 4.3 ± 0.57 | 0.54 | 60 |
| Spinal cord | 4.9 ± 0.31 | 4.7 ± 0.49 | 0.62 | 65 | 4.1 ± 0.6 | 3.9 ± 0.45 | 0.66 | 70 |
| Discs | 4.5 ± 0.69 | 4.7 ± 0.59 | 0.71 | 75 | 3.9 ± 0.51 | 3.9 ± 0.69 | 0.66 | 70 |

Table 4: Image quality assessment scores for the whole spine cases acquired using HASTE-VFA along with the inter-observer reliability values.

| Scoring Criteria | HASTE-VFA Mean Score | Gwet's AC1 | % Agreement |
|---|---|---|---|
| Motion | 4.7 ± 0.46 | 0.38 | 44.4444 |
| Edge sharpness | 4.2 ± 0.55 | 0.24 | 33.3333 |
| Artifacts | 4.4 ± 0.51 | -0.01 | 11.1111 |
| Noise | 4.4 ± 0.61 | -0.16 | 0 |
| Facet joints | 4.3 ± 0.49 | 0.5 | 55.5556 |
| Endplates | 4.7 ± 0.59 | 0.5 | 55.5556 |
| Spinal cord | 4.3 ± 0.69 | 0.35 | 44.4444 |
| Discs | 3.9 ± 0.42 | 0.64 | 66.6667 |